\documentclass[fleqn,usenatbib]{mnras}

\usepackage[T1]{fontenc}

\DeclareRobustCommand{\VAN}[3]{#2}
\let\VANthebibliography\thebibliography
\def\thebibliography{\DeclareRobustCommand{\VAN}[3]{##3}\VANthebibliography}

\usepackage{graphicx}	
\usepackage{amsmath}	
\usepackage{amssymb}	
\usepackage{multirow,subfigure}
\usepackage{xcolor}

\usepackage{newtxtext,newtxmath}

\title[Hanny's Voorwerp at 150\,MHz]{Relic jet activity in ``Hanny’s Voorwerp'' revealed by the LOFAR Two metre Sky Survey}

\author[D.J.B.~Smith et al.]{
D.~J.~B.~Smith,\thanks{E-mail: d.j.b.smith@herts.ac.uk}
M.~G.~Krause, M.~J.~Hardcastle and A.~B.~Drake
\\
Centre for Astrophysics Research, Department of Physics,
Astronomy and Mathematics, University of Hertfordshire, Hatfield AL10 9AB, UK
}

\date{Accepted XXX. Received YYY; in original form ZZZ}

\pubyear{2022}

\begin{document}
\label{firstpage}
\pagerange{\pageref{firstpage}--\pageref{lastpage}}
\maketitle

\begin{abstract}
We report new observations of ``Hanny's Voorwerp'' (hereafter HV) taken from the second data release of the LOFAR Two-metre Sky Survey (LoTSS). HV is a highly-ionised region in the environs of the galaxy IC2497, first discovered by the Galaxy Zoo project. The new 150\,MHz observations are considered in the context of existing multi-frequency radio data and archival narrow-band imaging from the \textit{Hubble Space Telescope}, centred on the \textsc{[Oiii]} emission line. The combined sensitivity and spatial resolution of the LoTSS data -- which far exceed what was previously available at radio frequencies -- reveal clear evidence for large-scale extended emission emanating from the nucleus of IC2497. The radio jet appears to have punched a hole in the neutral gas halo, in a region co-located with HV. The new 150\,MHz data, alongside newly-processed archival 1.64\,GHz eVLA data, reveal that the extended emission has a steep spectrum, implying an age $>10^8$\,yr. The jet supplying the extended 150\,MHz structure must have ``turned off'' long before the change in X-ray luminosity reported in recent works. In this picture, a combination of jet activity and the influence of the radiatively efficient active galactic nucleus are responsible for the unusual appearance of HV.
\end{abstract}




\section{Introduction}
\label{sec:intro}

Hanny's Voorwerp (hereafter HV) is a region of highly ionised material tens of kpc in projected size, which is located near the $z = 0.05$ galaxy IC2497. HV was originally reported in \citet{lintott2009}, having been discovered during visual classification of galaxies in the Sloan Digital Sky Survey \citep[SDSS;][]{york2000}, as part of the Galaxy Zoo project \citep{lintott2008}. HV was identified as a result of its unusual morphology and extremely bright {\sc{[Oiii]}} emission falling in the SDSS g' band filter. \citet{lintott2009} suggested that HV could be explained either as a highly photoionised region resulting from an active galactic nucleus (AGN) with an unusual dust geometry that prevents it from ionising the host galaxy's own nuclear gas, or as a ``light echo" resulting from a dramatic change in the luminosity of the central source over the past $10^5$ years.  

As well as discovering a halo containing $\sim 10^9$ M$_\odot$ of HI gas surrounding IC2497 and HV, \citet[hereafter J09]{jozsa2009} provided the first evidence of jets visible as a marginal extension to an already elliptical restoring beam in both 1.4 and 4.9\,GHz observations in the direction of HV. This extension is not visible in 1.4\,GHz observations from the Faint Images of the Radio Sky at Twenty-centimetres survey \citep[FIRST;][]{becker1995}, the NRAO VLA Sky Survey \citep[NVSS;][]{condon1998}, the Westerbork Northern Sky Survey \citep[WENSS;][]{rengelink1997}, or in 150\,MHz observations from the TIFR GMRT Sky Survey Alternative Data Release \citep[TGSS-ADR;][]{intema2017}. 
Nevertheless, \citet{rampdarath2010} further underlined the presence of jets using MERLIN observations to detect two components separated by $\sim$300\,pc with brightness temperatures in excess of $10^5$\,K \citep[an upper limit to the brightness temperature that can be attributed to star forming regions;][]{condon1991,biggs2010}, though with a significant flux deficit relative to the integrated measurements reported by J09 indicating the presence of a resolved nuclear starburst in IC2497. 

\textit{Hubble Space Telescope} observations in both broad- and narrow-band filters centred on the \textsc{[Oiii]} and H$\alpha$+\textsc{[Nii]} emission lines \citep{keel2012_hst} found evidence for large regions photoionised by AGN activity, with evidence for some sites within HV dominated by star formation. \citet{keel2012_hst} also hypothesised about the role of a previous major merger in stirring up the \textsc{Hi} gas around IC2497, and producing the bar and strong warping visible in the disk of IC2497. 

Several works have looked at HV at X-ray wavelengths, including \citet{schawinski2010} and \citet{sartori2018}. These works show that HV contains a Compton-thick AGN which has recently (in the last $\sim$70,000\,yr) undergone a dramatic change in luminosity, similar to the change suggested by \citet{keel2012a} and \citet{lintott2009}. Most recently, \citet{fabbiano2019} found evidence for extended soft X-ray emission in \textit{Chandra} data of HV, spatially consistent (albeit at low-resolution and with low statistical significance) with the direction of the extension similar to that in the radio data observed by J09. 

Our observational capabilities at radio frequencies have  exploded since HV was first studied with interferometry by J09 and \citet{rampdarath2010}. The capabilities of the Low Frequency Array \citep[LOFAR;][]{vanhaarlem2013} exemplify the huge steps in survey speed, sensitivity, resolution and imaging capabilities. One of the key drivers for LOFAR since its inception has been to conduct surveys of the whole northern sky, and the LOFAR Two-metre Sky Survey \citep[LoTSS][]{shimwell2017} is well on the way to fulfilling that aim. The first data release  \citep[LoTSS DR1;][]{duncan2019,shimwell2019,williams2019} covered 424\,deg$^2$ over the HETDEX \citep{hill2008} spring field using the LOFAR High Band Antenna (HBA) at a central frequency of 150\,MHz with a mean RMS sensitivity $< 100\,\mu$Jy and resolution around 6 arcsec. As well as the advantages of a huge increase in sensitivity relative to FIRST, LoTSS HBA data benefit from including short and long-baseline observations in the pipeline-processed data products, meaning that it is uniquely sensitive to low-frequency emission on both compact and extended scales.
LoTSS also includes even deeper observations over tens of deg$^2$ in the prime Northern fields with the best ancillary data (Bo\"otes, Lockman Hole and ELAIS-N1) in the first data release of the LoTSS Deep fields \citep{tasse2021,sabater2021,kondapally2021,duncan2021}.

Progress continues to be rapid; as well as observations at even lower frequencies \citep[using the Low Band Antenna;][]{degasperin2021} and using international baselines to obtain sub-arsecond resolution \citep{Morabito2022}, the second data release of LoTSS \citep[DR2;][]{shimwell2022} reaches a median 150\,MHz sensitivity of 84\,$\mu$Jy\,beam$^{-1}$ over $5700$\,deg$^2$, detects more than 4.3\,million 150\,MHz sources, and benefits from a range of further improvements to the data processing relative to LoTSS DR1.

In addition to the sensitivity of the LoTSS maps, together with higher frequency data, 150\,MHz observations are particularly useful for determining the properties of extragalactic radio sources. Energy loss in a population of relativistic electrons is expected to vary as a function of $\nu^2$, such that higher frequencies fade more rapidly as the electron population ages, and as a result the degree of radio spectral curvature increases with time. When studied in the context provided by GHz data therefore, low frequency observations can `anchor' the spectrum, and in doing so enable the best measurements of spectral curvature \citep[and therefore the age of the electron population; e.g.][]{kardashev1962,laing1980,harwood2013}. Using data spanning $325\,\mathrm{MHz} < \nu < 4.9\,$GHz, J09 reported a flat radio spectrum of the core of IC2497, consistent with a power law such that $S_\nu \propto \nu^{-\alpha}$ with an index of $\alpha = 0.55$, and no evidence for spectral curvature indicating that the core radio source is young. 

The IC2497-HV system is thought to be virtually unique in the local Universe \citep{keel2012a,keel2015}. Even in the absence of HV, the unambiguous presence of jets in IC2497 alone would make the system remarkable since \citet{singh2015} identified only four instances of double radio jets among 187,000 spiral galaxies identified in SDSS imaging  \citep[see also][]{mao2015,nesvadba2021}. HV therefore represents a prototype for studying a broad range of phenomena: the variability thought to be inherent in AGN physics, the role of mergers and\slash or interactions in triggering AGN activity, as well as the influence of AGN activity on the surrounding gas. These are some of the key physical processes required to explain the observed properties of galaxies \citep[see e.g.][]{silk1998,croton2006,alexander2012}. In this paper, we examine the properties of IC2497 and HV in the new 150\,MHz data from LoTSS DR2, and in newly processed archival 1.6\,GHz eVLA data. In Section \ref{sec:data} we describe the data we use, while in Section \ref{sec:results} we present our results before making some concluding remarks in Section \ref{sec:conclusions}.


\section{Data}
\label{sec:data}

\begin{figure*}
    \centering
    \subfigure{\includegraphics[width=0.9\columnwidth]{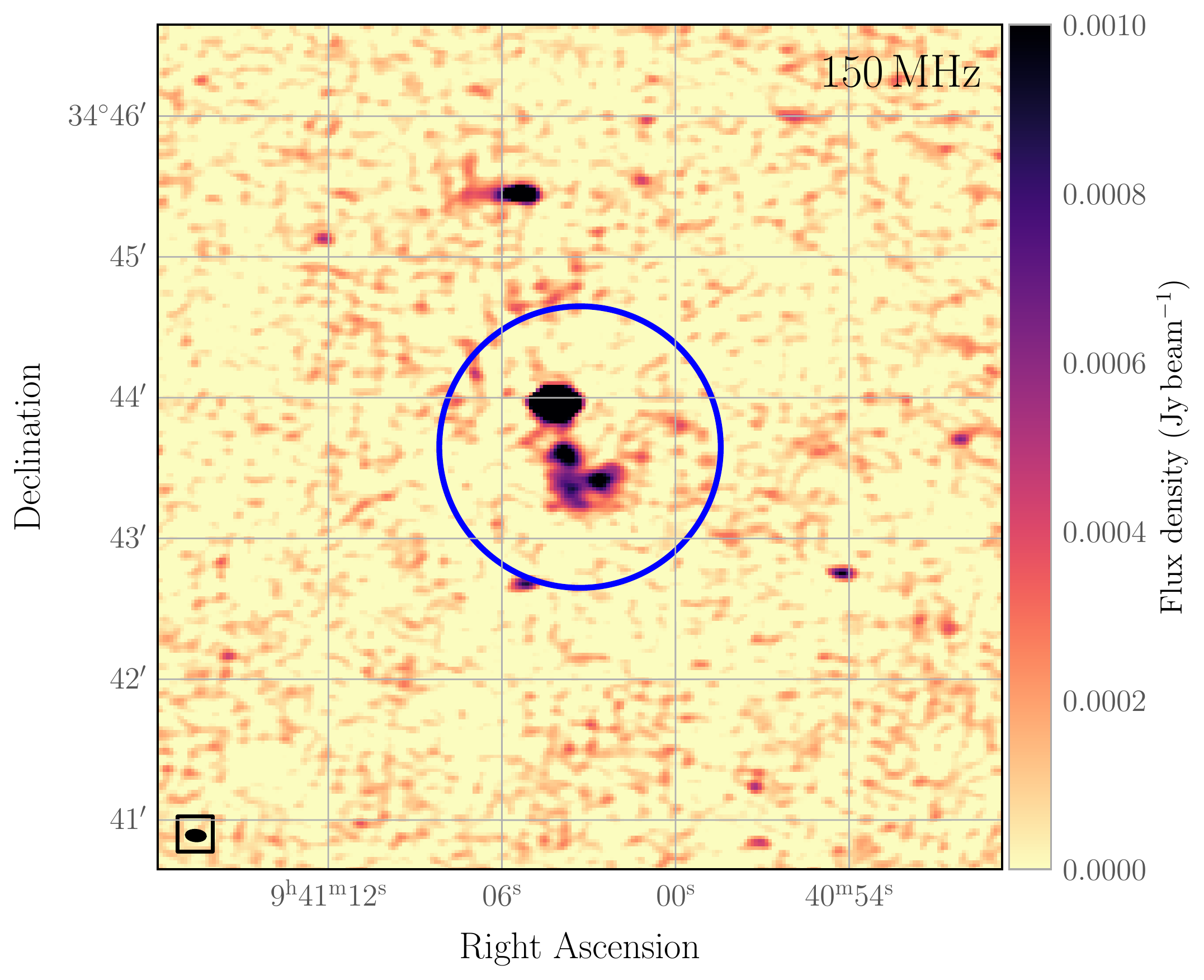}}
    \hspace{0.4cm}
    \subfigure{\includegraphics[width=0.9\columnwidth]{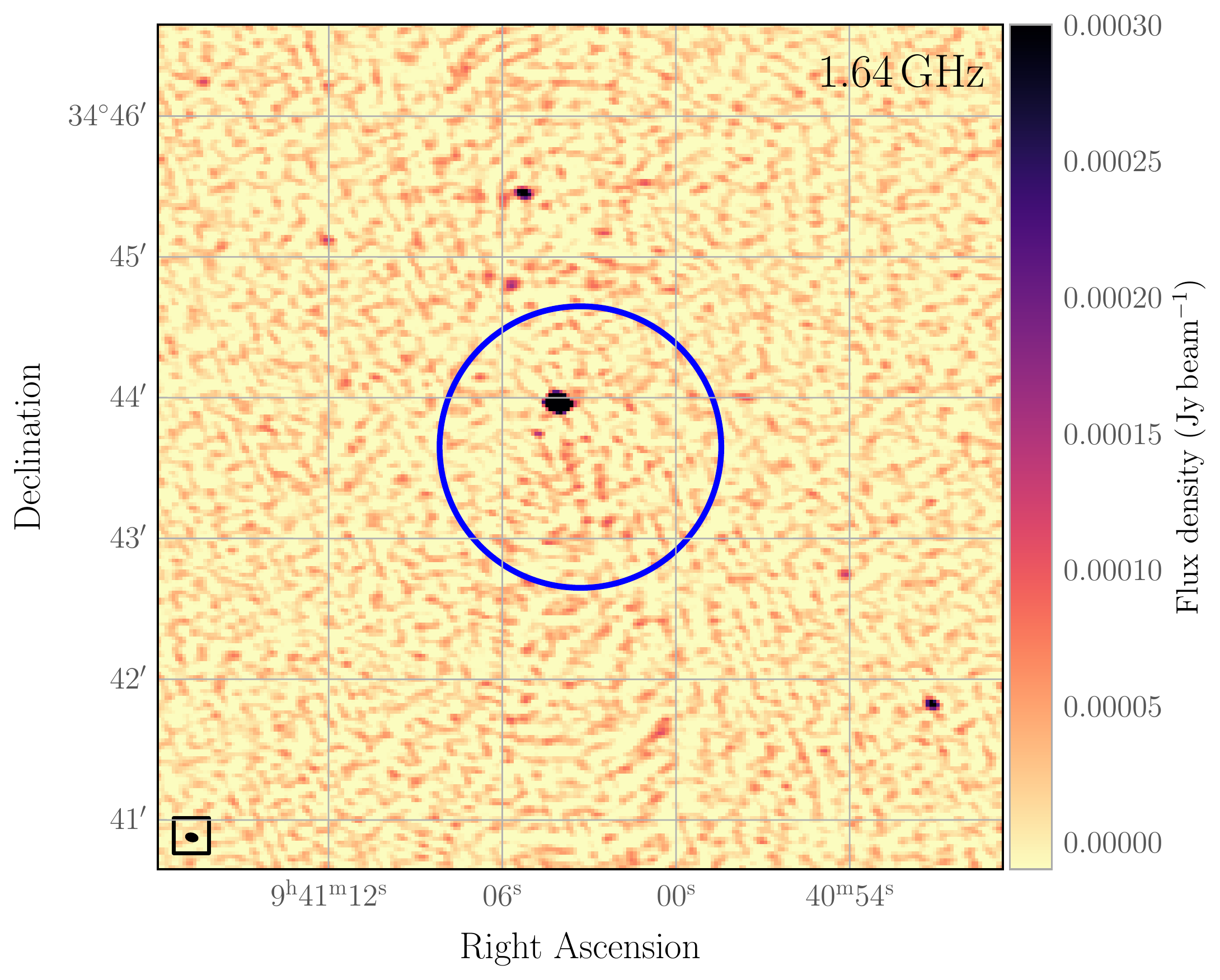}}
    \caption{New LoTSS (left) and eVLA (right) images of the region surrounding IC2497 and HV, highlighted with the blue circle, which has a radius of 1\,arcmin. The size of the natural-resolution restoring beam is shown as the black ellipse within a box in the lower-left corner of each image. The flux scale is as indicated by the colour bars to the right.}
    \label{fig:lotss_data}
\end{figure*}

New LoTSS DR2 observations of HV are available from the LOFAR surveys website\footnote{\url{www.lofar-surveys.org}}. The properties of the LoTSS DR2 mosaics are described in detail in \citet{shimwell2022}, however the vital statistics can be summarised as follows. The 150\,MHz images are automatically reduced using the latest version of the LoTSS processing pipeline, which uses direction-dependent methods to account for varying ionospheric conditions, giving high quality data with 6-arcsec resolution, and sensitivity to emission on both small and large scales due to the range of baselines available to the Dutch LOFAR array. HV falls within the area covered by the P144+35 mosaic; however, to obtain an optimal image of the region surrounding HV, we produced a self-calibrated image following the method of \cite{vanweeren2021}. The resulting image, shown in the left panel of Figure~\ref{fig:lotss_data}, is at the natural resolution of the LOFAR observations using robustness 0.5, and has an elliptical restoring beam with major and minor axes 8.5 and 4.8\,arcsec, 1.5\,arcsec pixels and a uniform RMS noise level of 92\,$\mu$Jy\,beam$^{-1}$ across the image. The absolute flux scale of the LoTSS data is thought to be correct at the level of 6\,percent.

We obtained archival eVLA data of the region surrounding HV, taken in
2011 in B-configuration (giving excellent sensitivity to extended
structures out to angular scales of 2\,arcmin). The eVLA data have an
effective frequency of 1.64\,GHz, with 256 MHz of bandwidth, and
provide similar spatial resolution (major and minor beam axes of 4.9 and 3.2\,arcsec) to the new LoTSS data. The newly-reduced eVLA map is shown in the right-hand panel of figure \ref{fig:lotss_data}; based on comparing the flux densities of unresolved sources with FIRST, we conclude that the flux scale is correct to $\sim5$\,percent.  We also obtained the 1.4\,GHz WSRT image and \textsc{Hi} data cube from J09 (G.I.G. J\'ozsa, private communication).

The HV-IC2497 system was observed using \textit{Advanced Camera for Surveys} (\textit{ACS}) tunable ramp filters on board the \textit{HST} \citep{keel2012_hst}, with the central wavelengths set to sample the redshifted \textsc{[Oiii]} and H$\alpha$ emission lines, and integration times of 2570\,s and 2750\,s respectively. 
The pointing was chosen so that HV and IC2497 fell within the monochromatic field of view (40\,arcsec $\times$ 80\,arcsec) and no continuum subtraction was attempted since \citet{lintott2009} showed that it was unnecessary. The final reduced HST data products from \citet{keel2012_hst} were provided by the authors (W.C. Keel, private communication).

\section{Results}
\label{sec:results}

\subsection{Morphology and photometry}

\begin{figure}
	\includegraphics[width=0.47\textwidth]{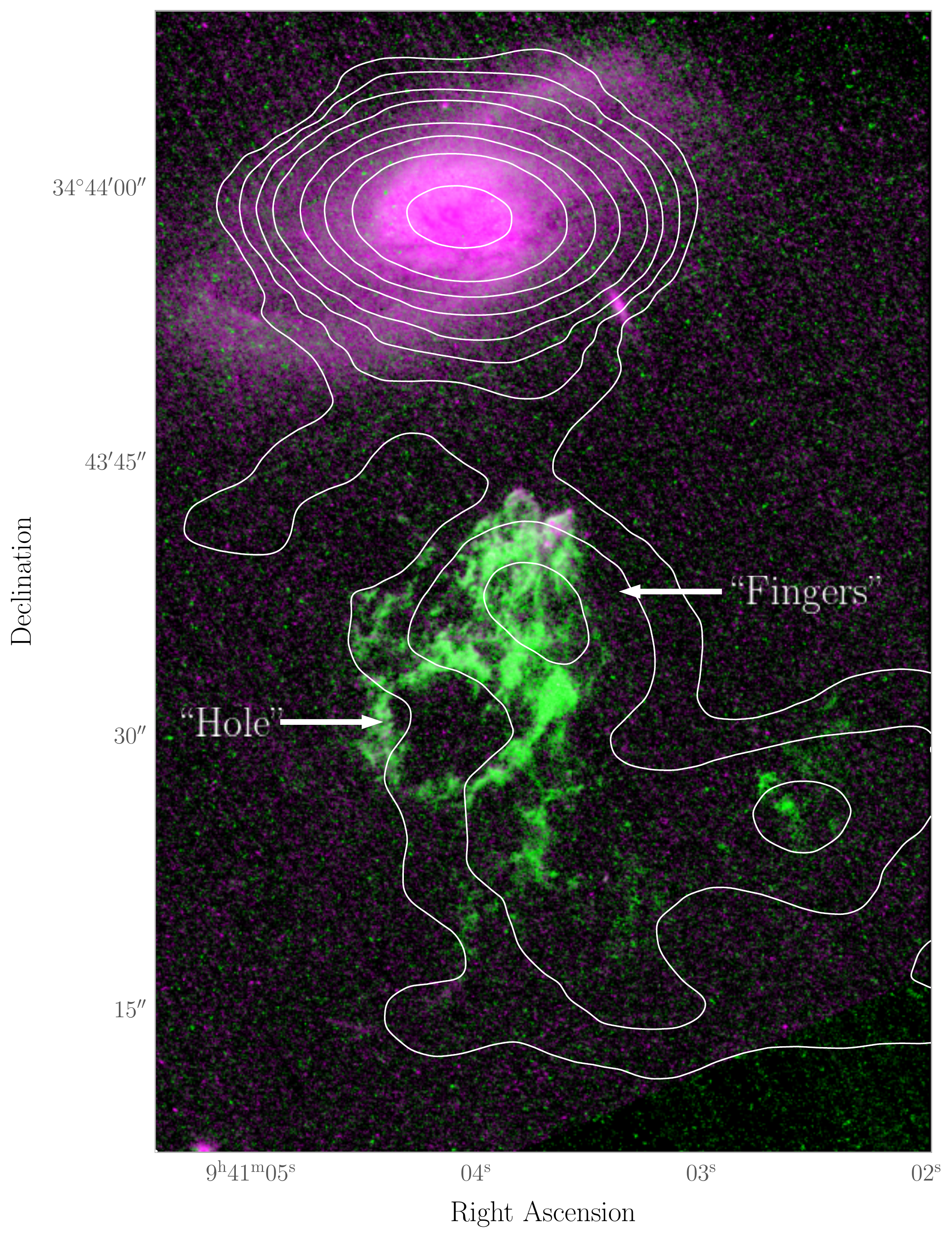}
    \caption{LoTSS 150\,MHz map of the region centred on HV (white contours) overlaid on a colour composite built from the [O\textsc{iii}] (green channel) and H$\alpha$ (blue and red channel) narrow-band filter HST images from \citet{keel2012_hst}. The LoTSS contours, shown in magenta, are at levels of 0.2, 0.5, 1, 2, 5, 10, 20 \&\ 50\,mJy\,beam$^{-1}$.}
    \label{fig:lotss_hst}
\end{figure}

Figure \ref{fig:lotss_hst} shows the new LoTSS 150\,MHz mosaic as
contours overlaid on a colour composite image derived using the
\textit{HST} \textsc{[Oiii]} and H$\alpha$ narrow-band images from
\citet{keel2012_hst}. It is clear that as well as an unresolved
component coincident with the centre of IC2497, the extension
identified by J09 is now resolved into clear jet-related emission spatially
coincident with HV, and with a flux density of $10.0 \pm 0.1$\,mJy.
The total flux density of the system at 150\,MHz in the new LoTSS mosaic is $97.3 \pm 0.2$\,mJy. This structure is not detected in the eVLA image (or any of the other assembled radio data), implying that its spectrum is steep.

The flux density of the unresolved core of IC2497 (obtained by subtracting $S_\nu^\mathrm{Extended}$ from  $S_\nu^\mathrm{Total}$) is consistent with the measurement from the TGSS-ADR mosaic once the calibration uncertainties are taken into account, and corresponds to a 150\,MHz luminosity of $(5.63 \pm 0.01) \times 10^{23}$\,W\,Hz$^{-1}$. Assuming a standard relationship from \citet{heckman2014} and correcting to 150\,MHz following \citet{sabater2019}, this enables us to estimate a mechanical power in the jet of $1.5\times 10^{36}$\,W. Although it compares well with the \citet{lintott2009} estimate of HV's \textsc{[Oiii]} luminosity ($1.5\times 10^{35}$\,W), the energetics alone preclude the current core activity from making the dominant contribution to the \citet{keel2012_hst} estimate of the luminosity required to maintain the ionization of HV ($\sim 10^{38}$\,W). 

Using the 100\,$\mu$m flux density quoted for IC2497 in the IRAS faint source catalogue \citep{moshir1990}, we can estimate the star formation rate (SFR) assuming that the dust is heated solely by young stars, that it has a far-infrared SED similar to M82 \citep{polletta2007} and the canonical relationship between far-infrared luminosity and SFR from \citet{kennicutt1998} adjusted to our adopted initial mass function (IMF) from \citet{chabrier2003}. Doing so, we obtain a far infrared luminosity integrated between 8-1000\,$\mu$m of $4.1\times 10^{11} L_\odot$ -- putting IC2497 in the Luminous Infrared Galaxy (LIRG) class -- and an SFR of $\sim 40\,$M$_\odot$\,yr$^{-1}$, although we note that the uncertainties inherent in our choice of SED template and IMF (and therefore the derived SFR) are significant. Nevertheless, comparing the derived SFR with the mass-independent $\mathrm{SFR}-L_{\mathrm{150\,MHz}}$ relations from \citet{gurkan2018} and \citet{smith2021}, reveals no evidence for any radio excess, consistent with the nuclear starburst reported by \citet{rampdarath2010}. Nevertheless, the morphological evidence for the presence of an AGN is conclusive.

\begin{table*}
    \centering
    \begin{tabular}{c|c|c|c|c}
      \hline
        Frequency & $S_\nu^\mathrm{Total}$ (mJy) & $S_\nu^\mathrm{Extended}$ (mJy) & Facility\slash Survey & Reference \\
        \hline 
        \multirow{2}{*}{150\,MHz\ \ {\bigg\{}}  & $97.3\pm0.2$ & $10.0 \pm 0.1$ & LoTSS &  This work  \\
          & $66.4\pm8.3$ & & TGSS-ADR &  \citet{intema2017}  \\  
        325\,MHz & $51 \pm 4.9$ & & WENSS & \citet{rengelink1997} \\
        \multirow{3}{*}{1.4\,GHz\ \ \  {\Bigg\{}} & $20.9 \pm 1.1$ & $3.2\pm 0.2$ & WSRT & \citet{jozsa2009} \\
         & $18.5 \pm 0.6$ & & NVSS & \citet{condon1998} \\
         & $16.8 \pm 0.9$ & & FIRST &  \citet{becker1995} \\
        1.64\,GHz & $14.2 \pm 0.5$ & $0.48 \pm 0.29$ & eVLA & This work \\
        4.9\,GHz & $11.6 \pm 0.6$ & & WSRT & \citet{jozsa2009}\\
        \hline
    \end{tabular}
    \caption{A compilation of radio frequency flux densities of IC2497 \&\ HV, including both the total ($S_\nu^\mathrm{Total}$) and extended ($S_\nu^\mathrm{Extended}$) components where available.}
    \label{tab:fluxes}
\end{table*}

\begin{figure*}
	\subfigure{\includegraphics[height=0.89\columnwidth]{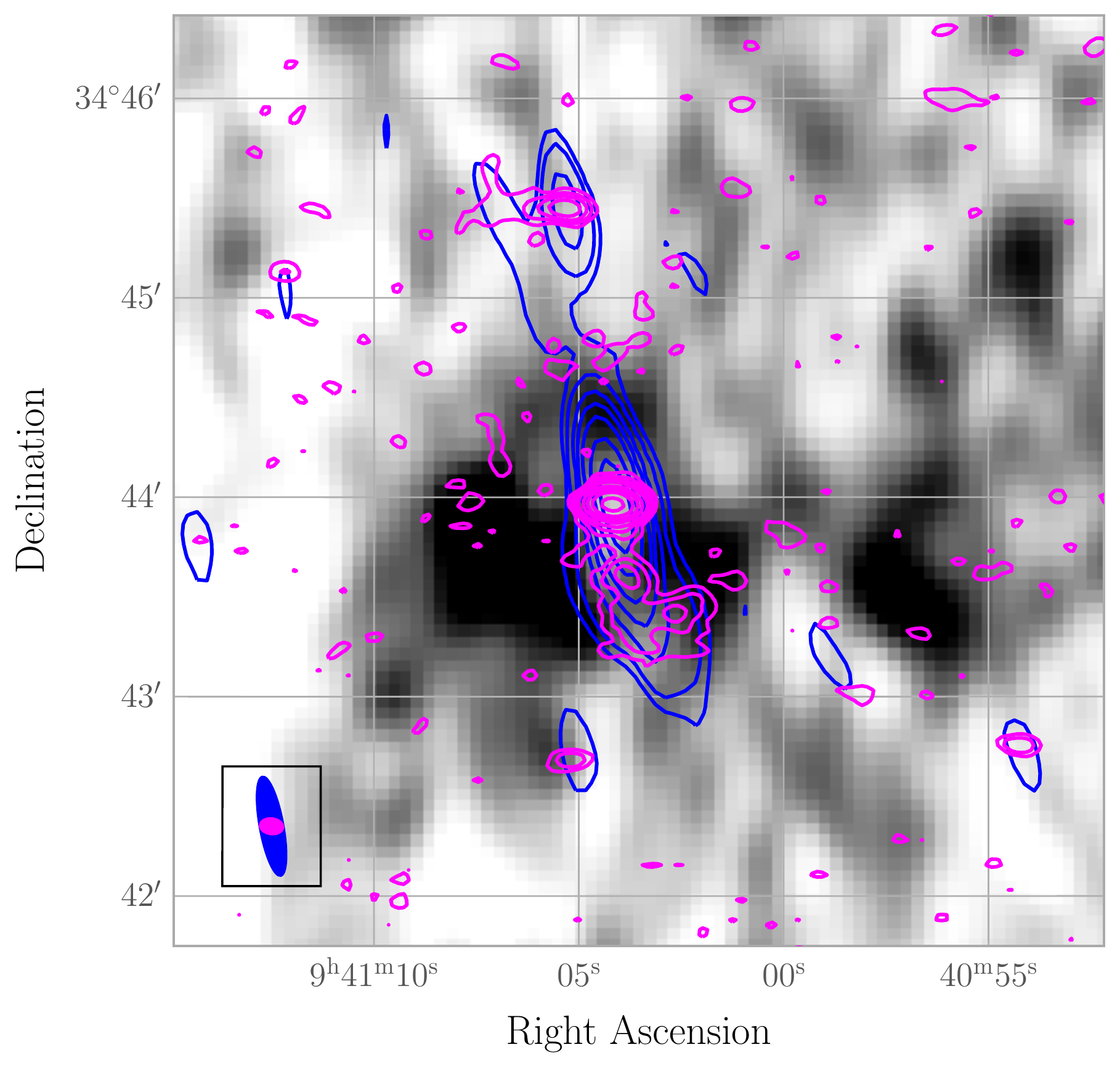}}
	\subfigure{\includegraphics[height=0.89\columnwidth]{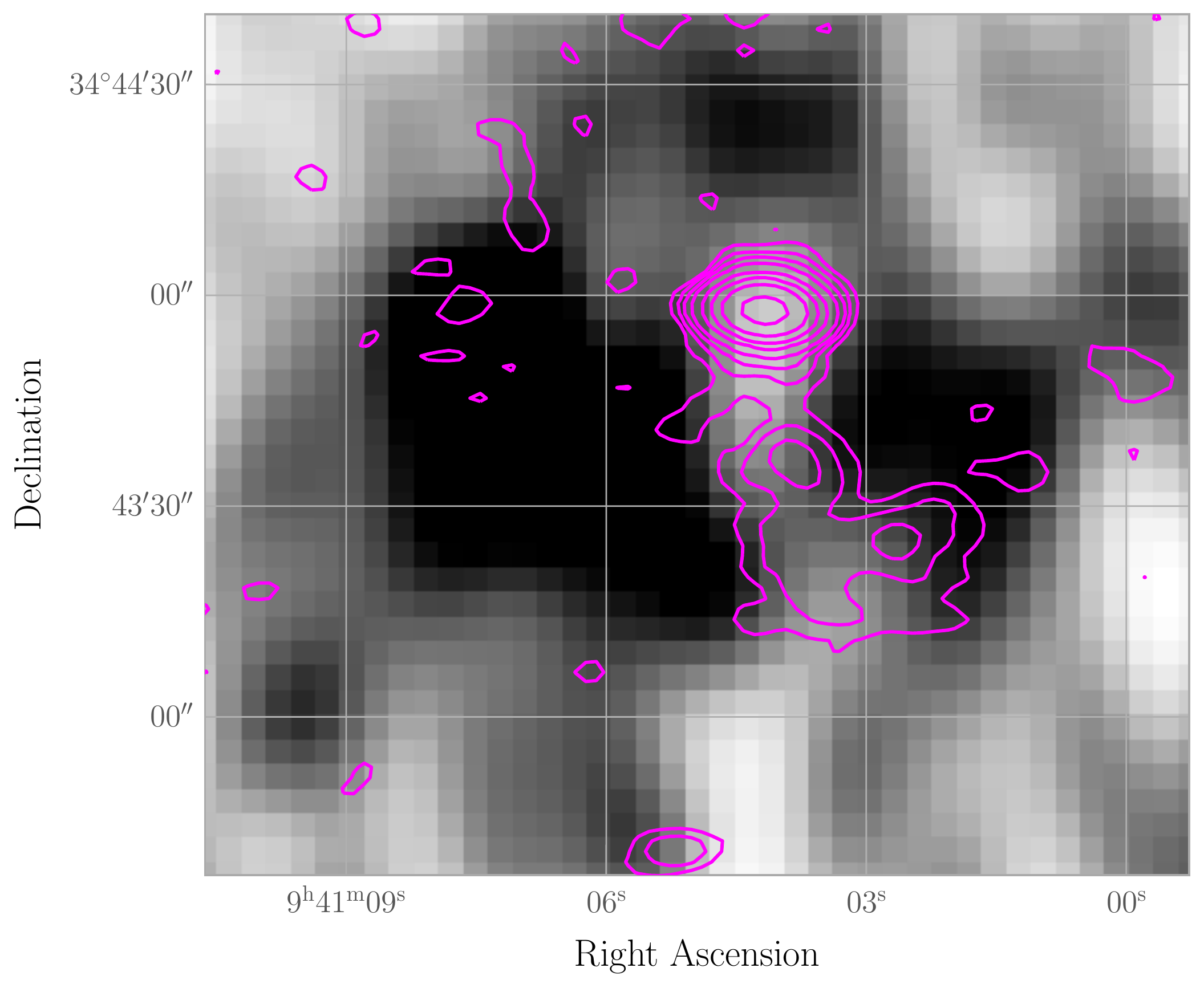}}
    \caption{Composite image showing the \textsc{Hi} image (background, greyscale) and the new LoTSS data (magenta contours, levels of 0.2, 0.5, 1, 2, 5, 10, 20 \&\ 50\,mJy\,beam$^{-1}$). In the left panel, the WSRT 1.4\,GHz image from \citet{jozsa2009} is shown as blue contours (with contour levels of 0.1, 0.2, 0.5, 1, 2, 5, 10 \&\ 20\,mJy\,beam$^{-1}$), and the filled ellipses in the box to the bottom left indicate the size of the restoring beam, colour-coded to match the contours. The right panel is zoomed in to highlight the spatial anti-correlation between the jet (magenta contours) and the \textsc{Hi} gas (background greyscale) with the same contour levels and colour scale as the left panel.}
    \label{fig:radiomaps}
\end{figure*}

To put the new morphological information available from the $\sim$6 arcsec
resolution LoTSS data in the context of the previously available radio
data, in both panels of Figure~\ref{fig:radiomaps} we overlay the 150\,MHz data as
magenta contours on a background image showing the \textsc{Hi} data
from J09. Overlaid in the left panel is the 1.4\,GHz WSRT image from J09, shown as
blue contours \citep[the WSRT data have an elliptical point spread
  function of $14 \times 11$ arcsec, shown as the blue-filled ellipse in the lower-left corner;][]{morganti2004}, with the
details of the chosen contour levels given in the caption. While at
1.4\,GHz the degree of extension in the radio source is perhaps
debatable on the basis of visual inspection alone due to the elongated
beam along the IC2497-HV direction, this is not true at 150\,MHz,
where the LoTSS PSF (shown as the magenta ellipse to the lower-left) is more than four times smaller. In the right panel of Figure~\ref{fig:radiomaps} we show a zoomed version of the central region; for the first time it is clear that the clearly-resolved structure emanating from the nucleus of IC2497 coincides with a minimum in the surrounding $10^9\,M_\odot$ reservoir of \textsc{Hi} gas found by J09 (shown in the background image).

Based on the left panel of Figure~\ref{fig:radiomaps} it would be tempting to identify the knot of coincident 150\,MHz and 1.4\,GHz emission on the Northern side of IC2497 as being related to the counter-jet, emanating in the direction opposite to HV. However, inspection of the SDSS images shows that this knot of emission is instead associated with an edge-on late-type galaxy interloper. 

\subsection{Radio spectral index}
\label{sec:alpha}

\begin{figure}
    \centering	\includegraphics[width=0.9\columnwidth]{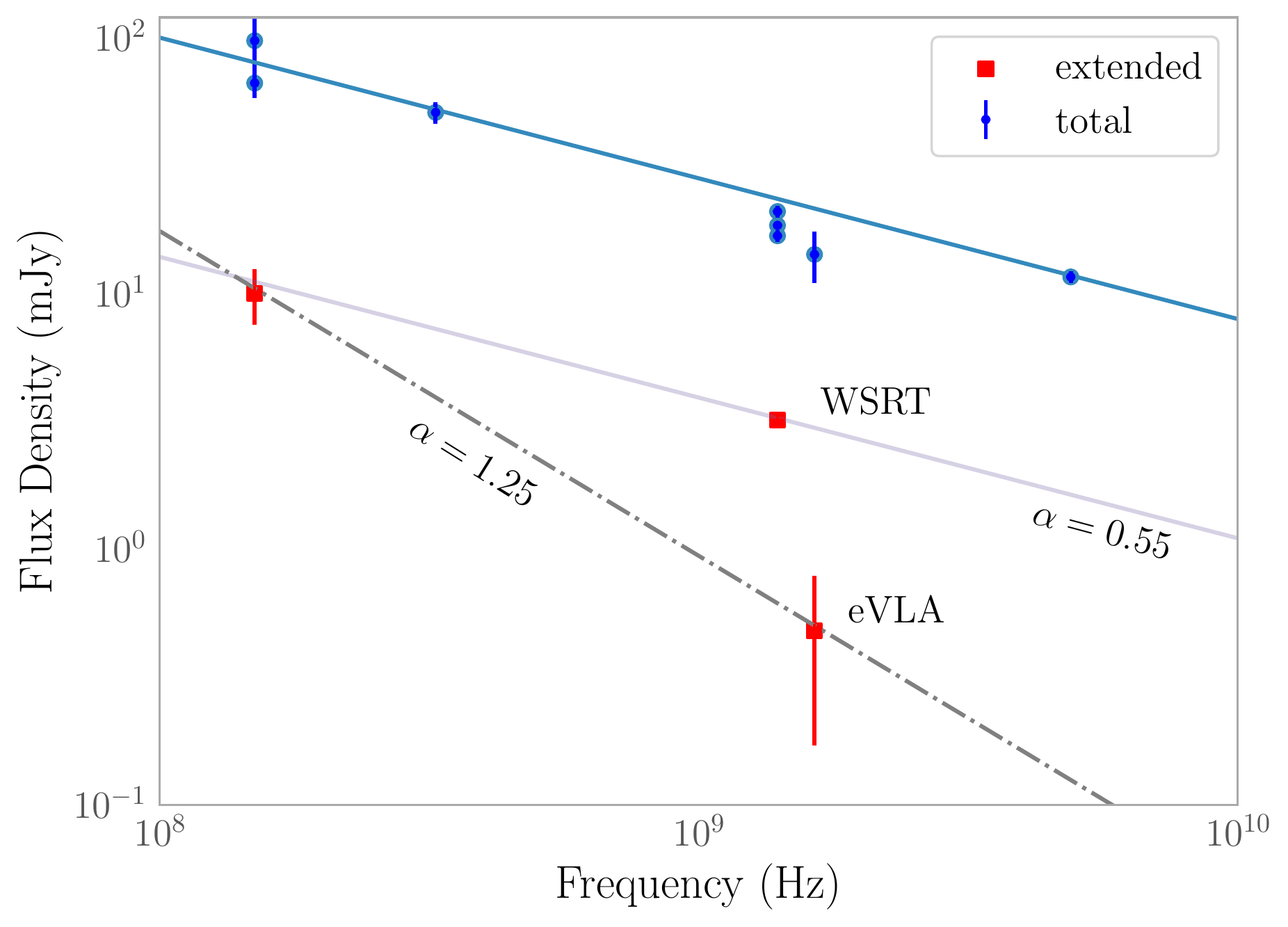}
    \caption{Radio frequency spectra of HV, including values for the total (blue circles) and extended components (red squares), as detailed in Table \ref{tab:fluxes}. The solid blue line indicates a power-law spectrum with spectral index $\alpha = -0.55$, while the dotted and dot-dashed lines indicate the spectral index obtained using the 150\,MHz data alongside the WSRT 1.4\,GHz or eVLA 1.64\,GHz flux densities, which are formally inconsistent as discussed in the text.}
    \label{fig:spectrum}
\end{figure}

To examine the radio spectrum of the IC2497-HV system, in Figure \ref{fig:spectrum} we show the total photometry ($S_\nu^\mathrm{Total}$) from Table \ref{tab:fluxes} as filled circles, though we have added calibration uncertainties in quadrature, at the level of 6\,percent to the new LoTSS flux density \citep{shimwell2022}, and 5\,percent to the new eVLA measurements. The photometry is overlaid with a power law with the spectral index $\alpha = 0.55$ found by J09. Despite reaching lower frequencies, the new LoTSS data do not reveal any evidence for spectral curvature in the core; the power law found by J09 continues down to frequencies of 150\,MHz at least. 

In figure \ref{fig:spectrum}, the extended source photometry
($S_\nu^\mathrm{Extended}$ from Table \ref{tab:fluxes}) is shown using square symbols.
There are conflicting results for the spectral index of the extended
component, depending on whether we use the WSRT or eVLA flux
densities, since the nominal 1.64\,GHz eVLA flux density of $0.48 \pm
0.29$\,mJy (measured at a spatial resolution similar to the new LoTSS
data) is inconsistent with the 1.4\,GHz WSRT estimate of $3.2\pm
0.2$\,mJy. We suggest that the flat spectrum of the extended emission ($\alpha_{150}^{1400} =
0.51 \pm 0.03$) obtained using the WSRT estimated flux density may be
an artefact of contamination from the extension in the North-South
direction of the WRST point spread function. Using instead the eVLA
flux density alongside the 150\,MHz measurement we obtain our best
estimate of the extended component's spectral index of
$\alpha_{150}^{1640} = 1.25^{+0.28}_{-0.23}$. As the extended
structure detected in the LOFAR data is not formally detected in the
eVLA image, the spectrum might be even steeper, but we can be confident
that $\alpha > 1$. 

The steep spectrum implies that the plasma in the extended structure
is significantly older than that in the core, as expected if the jet
feeding it has `turned off' some time ago. However, since we have measurements at just
two frequencies, and a detection only at 150 MHz, detailed analysis is
not possible. If we assume that the injection index for this material
is $\alpha = 0.55$, estimate a self-consistent equipartition magnetic
field strength of around 0.2 nT (based on treating the extended
material as a uniformly filled sphere of radius 16 arcsec), and take
account of inverse-Compton losses to the CMB at the redshift of
HV\footnote{Calculations were done using the {\sc pysynch} Python
package \url{https://github.com/mhardcastle/pysynch} which provides an
interface to the synchrotron emission code of \cite{hardcastle1998}.},
then using a Jaffe-Perola model for spectral ageing
\citep{jaffe1973} we find that the spectral age must be $\ga 10^8$ yr to give
$\alpha > 1$. Adiabatic expansion of a lobe generated by a now
disconnected jet will tend to increase the apparent spectral age,
perhaps by an order of magnitude, but this still implies that the jet
supplying the extended structure must have turned off long before the
change in X-ray luminosity reported by \citet{sartori2018}. Future
60\,MHz observations with the LOFAR Low-band Antennae may be able to
provide better constraints on the degree of spectral curvature, and
therefore on the age of the plasma.

\section{Discussion \&\ Conclusions}
\label{sec:conclusions}

We attempt to bring all of this information together as follows. The flat spectral index in the core of IC2497 extends to 150\,MHz, and is similar to values for injection indices assumed by jet models \citep[see e.g.][for a discussion]{shulevski2015}. This is consistent with the small-scale nuclear jets observed by \citet{rampdarath2010}.

On the other hand, the extended component seen at 150 MHz has a steep spectrum ($\alpha_{150}^{1640} = 1.25^{+0.28}_{-0.23}$), implying an age of order $10^8$ yr if the injection index is the same as that measured for the compact component. Other studies based at X-ray wavelengths \citep[e.g.][]{schawinski2010,sartori2018} have suggested a timescale of $\sim 10^5$ years for a change in the luminosity of the central engine; however the LoTSS data conclusively show that the jet that generated the extended emission must have been unrelated to that nuclear outburst.

The brightest knots of 150\,MHz emission visible in Figure~\ref{fig:lotss_hst} appear co-located with the main star-forming structure at the Northern end of HV, and with the apparently separate (in the \textit{HST} data) component to the West. The Northern 150\,MHz knot appears coincident with the filamentary ``fingers" located just to the South of the star-forming region located at the Northern tip of HV (which appears pink in figure \ref{fig:lotss_hst} due to the bright H$\alpha$ emission). These features were first identified in \citet{keel2012_hst} as possible evidence of gas entrained by ram pressure of the jet emanating from the core of IC2497. 
Radiative dissipation in turbulent mixing may contribute to the line emission locally \citep{KA07}, even though the
overall spectrum is dominated by photoionisation (compare below).

The ``hole" feature on the South-East side of HV is reminiscent of supernova remnants visible within our own galaxy, although the overall structure and the optical line ratios in this area measured by \citet[based on long-slit spectroscopy]{lintott2009} and \citet[based on \textit{HST} narrow-band imaging]{keel2012_hst} are clearly dominated by photoionisation, and therefore not associated with the passage of the jet. \citet{fabbiano2019} reported extended X-ray emission apparent in this area (albeit at low statistical significance) as might also be expected under the hypothesis of a recent explosive event at this location, although no remnant is visible in any of the assembled data.

In the \citet{krause2005} model, the leading shell of a starburst wind \citep[perhaps resulting from the nuclear starburst suggested to be present in the HV system by][]{rampdarath2010} can cool and form a dense, mainly neutral \textsc{Hi} shell, if for some reason it encounters enough gas, such as in the post-major merger scenario for IC2497 discussed by \citet{keel2012_hst}. When a large-scale jet is also present (as is clearly the case based on the new 150 MHz data), it interacts with the shell and pierces a hole in it. In the \citet{krause2005} simulations, the hole is $\sim$20\,kpc in diameter, very similar to the radio lobe width observed in HV.

A counterargument to this picture appears to be that in the \citet{lintott2009} optical spectroscopy, as well as in unpublished integral-field spectroscopy of the HV system using the WIYN-HEXPAK instrument (W.C. Keel, \textit{private communication}) HV has a smooth velocity field showing no evidence of disturbances that might be expected in a jet interaction scenario (see \citealt{krause2005}, Fig. 3). 
While a blue-shifted velocity field that is coherent over many kpc might indeed point to a starburst wind shell, the scenario from \citet{krause2005} could still describe HV but must be modified. An element of physics that was missing in the latter simulations was the Vishniac shell instability \citep{vishniac1983}: 3D high-resolution simulations have shown that such shells fragment into filaments and clumps \citep[e.g.][]{vanmarle2012,krause2013}. The disturbed kinematics in the \citet{krause2005} simulations come from the strong early resistance of the unfragmented shell with high pressure build-up inside of it and strong gas acceleration when the shell eventually fragments due to other instabilities. If the shell in reality fragments earlier, the jet might even better pierce the immediate impact region between the clumps with less effect on the surrounding shell. This could be tested with dedicated simulations.

Alternatively, the spatial coincidence between jet and \textsc{Hi} minimum could be purely a projection effect (in which case the \textsc{Hi} minimum may be apparent due to the gas at this location being almost fully ionised), or that the jet punched through the gas sufficiently long ago that any turbulence has subsided \citep[e.g.][]{krumholz2006}. If we assume that the timescale for turbulent decay is similar to the crossing time of the emission-line region, back of the envelope calculations (dividing the size of the region by the assumed speed, adopting an initial turbulent velocity of 100\,km\,s$^{-1}$, similar to velocities found e.g. from the simulation of \citealt{krause2005} or the observations of \citealt{nesvadba2021}, and a region size of 10\,kpc) produce timescales of the order of $10^8$\,yr, which is comparable to the lower limit on the  lobe age estimated in section \ref{sec:alpha}. We would therefore expect any turbulent motions of this magnitude driven into the emission-line region by the expanding lobes not to be visible by the present time. 

The picture that emerges from the assembled data is one with multiple events over different timescales. First, a tidal encounter left a $10^9$\,M$_\odot$ cloud of \textsc{Hi} around IC2497, and caused IC2497's warped appearance. The new radio frequency data then show that a radio outburst, $\sim 10^8$ years ago, shaped and perturbed the
multi-phase gas 
tens of kpc around IC2497, before switching off. The relic radio lobe from this outburst is what we now
see in the LOFAR data -- a corresponding northern radio lobe has
presumably faded to invisibility, perhaps because of the lower gas
density to the north of IC2497.  Much more recently ($\sim 10^5$ years
ago) a radiatively-efficient AGN outburst illuminated the gas that 
had been swept up around the radio lobe and gave rise to the extended
emission-line region that characterizes HV. The ongoing nuclear starburst and jet \citep[both first observed by][]{rampdarath2010} may have
begun\slash turned on around the same time but it seems unlikely that the jet was continuously active through the period. This timeline is clearly in contrast with the previous suggestion that fading AGN could indicate a lasting change in accretion mode \citep[from being dominated by radiative to mechanical output; e.g.][]{sartori2018}. In the IC2497-HV system, the situation is clearly more complex, with evidence for recurring episodes of mechanical (or radiatively inefficient) jet activity in addition to the radiatively efficient activity responsible for ionising HV.

We echo previous suggestions that HV could be a unique local object that exemplifies processes much more common in the high-redshift Universe at the peak of cosmic star formation, and is therefore worthy of more study.


\section*{Acknowledgements}

DJBS dedicates this work to the loving memory of D.H.F. Smith (1943-2021): father, Chieftain, poet, and Voorwerp enthusiast. The authors would like to thank W.C. Keel for refereeing this paper, for providing the fully-processed \textit{HST} tunable ramp filter data from \citet{keel2012_hst}, and for a preview of WIYN-HEXPAK observations of HV. We also thank G.I.G. J\'osza for providing the FITS data products from \citet{jozsa2009}. The authors would like to thank Paul Haskell and Soumyadeep Das for valuable comments. DJBS and MJH acknowledge support from the UK Science and Technology Facilities Council (STFC) under grant ST/V000624/1. 
LOFAR is the Low Frequency Array designed and constructed by ASTRON. It has observing, data processing, and data storage facilities in several countries, which are owned by various parties (each with their own funding sources), and that are collectively operated by the ILT foundation under a joint scientific policy. The ILT resources have benefited from the following recent major funding sources: CNRS-INSU, Observatoire de Paris and Université d'Orléans, France; BMBF, MIWF-NRW, MPG, Germany; Science Foundation Ireland (SFI), Department of Business, Enterprise and Innovation (DBEI), Ireland; NWO, The Netherlands; The Science and Technology Facilities Council, UK; Ministry of Science and Higher Education, Poland; The Istituto Nazionale di Astrofisica (INAF), Italy. 

This research made use of the Dutch national e-infrastructure with support of the SURF Cooperative (e-infra 180169) and the LOFAR e-infra group. The Jülich LOFAR Long Term Archive and the German LOFAR network are both coordinated and operated by the Jülich Supercomputing Centre (JSC), and computing resources on the supercomputer JUWELS at JSC were provided by the Gauss Centre for Supercomputing e.V. (grant CHTB00) through the John von Neumann Institute for Computing (NIC). 

This research made use of the University of Hertfordshire high-performance computing facility and the LOFAR-UK computing facility located at the University of Hertfordshire and supported by STFC [ST/V002414/1], and of the Italian LOFAR IT computing infrastructure supported and operated by INAF, and by the Physics Department of Turin University (under an agreement with Consorzio Interuniversitario per la Fisica Spaziale) at the C3S Supercomputing Centre, Italy.

\section*{Data Availability}

The new LoTSS 150\,MHz observations presented in this work are available from the LOFAR Surveys website, \url{https://www.lofar-surveys.org/} as part of the second data release of the LOFAR Two-metre Sky Survey (LoTSS). 



\bibliographystyle{mnras}
\bibliography{refs} 




\appendix




\bsp	
\label{lastpage}
\end{document}